\documentclass[12pt]{article}
\begin{document}
\newcommand{\co}{\; \; ,}
\newcommand{\nn}{\nonumber \\}
\newcommand{\scs}{\co \;}
\newcommand{\per}{ \; .}
\newcommand{\unith}{{\bf{\mbox{1}}}}\newcommand{\sem}{\;\; ; \; }
\newcommand{\bpi}{{\mbox{\boldmath{$\pi$}}}}
\newcommand{\calleff}{{\cal L}_{\mbox{\small{eff}}}}
\newcommand{\call}{${\cal L}$}
\newcommand{\bea}{\begin{eqnarray}}
\newcommand{\eea}{\end{eqnarray}}
\newcommand{\be}{\begin{equation}}
\newcommand{\ee}{\end{equation}}

\thispagestyle{empty}
\begin{titlepage}
\begin{flushright}
BUTP--99/12\\
\end{flushright}
\vspace{1cm}
\begin{center}
{{\Large{\bf{{Chiral Dynamics}}\footnote{Invited talk given at the
International Workshop ``~$e^+e^-$ collisions from $\phi$ to
$J/\psi$~'', March 1-5, 1999, Budker Institute of Nuclear Physics,
Novosibirsk.}$^,$\footnote{
Work  supported in part by the Swiss National Science
Foundation, and by TMR, BBW--Contract No. 97.0131  and  EC--Contract
No. ERBFMRX--CT980169 (EURODA$\Phi$NE).
}
 }}}

\vspace{1cm}

 J.~Gasser\\[1cm]
 Institut f\"ur  Theoretische Physik, Universit\"at  Bern,\\
Sidlerstrasse 5, CH--3012 Bern, Schweiz \\
gasser@itp.unibe.ch\\[.5cm]
June 1999 \\[1cm]

{\bf{Pacs:}} 11.30.Rd, 13.20.Eb, 13.75.Lb

{\bf{Keywords:}}
 Chiral symmetries, semileptonic decays of $K$ mesons,
meson-meson interactions, pion-pion scattering

\end{center}

\vskip.5cm

\begin{abstract}
\noindent
After  a short summary of my talk, I discuss $K_{l3}$ decays and
elastic $\pi\pi$ scattering in the framework of chiral perturbation
theory.

\end{abstract}

\end{titlepage}

\clearpage

\setcounter{page}{2}
\section{Introduction}
In the first part of may talk, I gave an introduction to
 the effective theory of QCD at low energy, called  chiral
perturbation theory (CHPT) \cite{weinberg79,glannp}. There are many
excellent reviews and
lectures on the subject available on the market - for a comprehensive list,
 see Ref.~\cite{honnef}. Therefore, I do
not try to add  one more here. Instead, I refer the
interested reader to Refs. \cite{honnef} and \cite{granada}.
 In the second part, I illustrated the method with a few
 examples. Here,  I
 shall consider two of them, $K_{l3}$ decays and elastic
$\pi\pi$ scattering. Both processes are presently  under theoretical and
experimental investigation.
Finally, I also presented  the EURODAFNE network, outlining
 the work planned in that enterprise. Lack of space prevents  me to
cover this topic here - I refer the interested reader to  the
 relevant homepages \cite{homepageeurodafne} and to
{\em The Second DAFNE Physics Handbook}
 \cite{handbook}.

\section{$K_{l3}$ decays}

The so called $K_{l3}$ decays are
\bea
 K^+(p) &\rightarrow& \pi^0 (p') l^+ (p_l) \nu _l (p_\nu) \hspace{1cm}
[K_{l3}^+] \label{s31}\nonumber
\\
K^0(p) &\rightarrow &\pi^- (p') l^+ (p_l) \nu_l (p_\nu) \hspace{1cm} [K_{l3}^0]
\label{s32}
\eea
and their charge conjugate modes. The symbol  $l$ stands for $\mu$ or
$e$. I consider the isospin symmetry limit $m_u=m_d,
\alpha_{\mbox{\tiny{QED}}}=0$.

The matrix element for $K_{l3}$ decays contains a leptonic and  a
hadronic factor. The hadronic part is
\bea
\hspace{-.8cm}\langle\pi^0 (p') \mid V_\mu^{4-i5} (0)\mid K^+(p)\rangle
=&&
\langle\pi^-(p') \mid V_\mu^{4-i5} (0)\mid K^0(p)\rangle\nonumber\\
=&& \frac{1}{\sqrt{2}} [(p'+p)_\mu f_+ (t) + (p-p')_\mu
f_- (t)]\, .
\eea
In this formula, $V_\mu^{4-i5}$ denotes
the hadronic vector current, and $t$ is the
momentum transfer to the lepton pair, $ t=(p'-p)^2 = (p_l+p_\nu)^2$.

The quantity $f_+$ is referred to as the vector form factor, because
it specifies the P-wave  projection of the crossed channel matrix element.
% $\langle 0 \mid V^{4-i5}_\mu(0) \mid K^+, \pi^0 \;\mbox{in} \rangle$.
 The S-wave projection is described by the scalar form factor
\bea
f_0 (t) = f_+ (t) + \frac{t}{M^2_K - M^2_\pi} f_-(t) \; \;
{}.
\eea
Analyses of $K_{l3}$ data often assume a linear dependence
\bea
f_{+,0} (t) = f_+ (0) \left[ 1 + \lambda_{+,0}
\frac{t}{M^2_{\pi^+}} \right] \; \; .
\eea

\subsection{Previous measurements}

I refer the reader to the 1982 version of the PDG \cite{pdg82}
for a critical discussion of the wealth of experimental information on
$\lambda_{+,0}$. Here I present  a short summary.

\underline{$K_{e3}$-experiments}

The $\lambda_+$ values obtained are fairly consistent. The average values are
\bea
K^+_{e3} \;:\; \lambda_+ &=& 0.0286 \pm 0.0022 \; \; \; \;
 \cite{pdg}
 \nonumber\\
K_{e3}^0 \ : \, \lambda_+& =& 0.0300 \pm 0.0016 \; \; \;\; \cite{pdg} \;
\; .
\label{s318}
\eea

\underline{$K_{\mu 3}$-experiments}

The result by Donaldson et al. \cite{donaldson}
\bea
\lambda_+& =& 0.030 \pm 0.003
\nonumber \\
\lambda_0& =& 0.019 \pm 0.004
\label{s319}
\eea
dominates  the statistics in the $K^0_{\mu3}$ case. The $\lambda_+$ value
(\ref{s319}) is consistent with the $K_{e3}$ value (\ref{s318}).
 However, the situation concerning the slope $\lambda_0$
is rather unsatisfactory, as the following  list
from $K_{\mu 3}^0$ decays illustrates\footnote{The list is
chronological, starting 1974, ending 1981. Earlier data
may be found in Ref.~\cite{pdg}. More recent data are not yet available.}
\be
\lambda_0 = \left\{\begin{array}{llll}
              0.019 &\pm &0.004  & \cite{donaldson}\\
              0.025 &\pm &0.019  & \cite{buchanan} \\
              0.047 &\pm &0.009  & \cite{clark} \\
              0.039 &\pm &0.010  &  \cite{hill} \\
              0.050 &\pm &0.008  & \cite{cho} \\
              0.0341 &\pm &0.0067 & \cite{birulev} \; .

                \end{array}
              \right.
\label{l:s324}
\ee
The $\chi^2$ fit to the $K^0_{\mu3}$ data yields $\lambda_+ = 0.034 \pm 0.005$,
$\lambda_0 = 0.025 \pm 0.006$ with a $\chi^2/DF = 88/16$ \cite[p.76]{pdg82}!
The situation in the charged mode $K^+_{\mu3}$ is slightly better \cite{pdg82}.

\subsection{Theory}

The theoretical prediction of $K_{l3}$ form factors has a long history,
starting
in the sixties with the current algebra evaluation of $f_{\pm,0}$. For
an
early review of the subject and for references to work prior to CHPT
evaluations
of $f_{\pm,0}$, I refer the reader to \cite{chounet}.
 Here I concentrate on the
evaluation of the form factors in the framework of CHPT.
The one-loop corrections have  been evaluated in \cite{glnpform},
with the result
\bea
\lambda_0 &=& 0.017 \pm
0.004\, ,
\label{s333}
\eea
where the error is an estimate of the uncertainties due to higher-order
contributions. The prediction (\ref{s333}) is in agreement with the
high-statistics experiment \cite{donaldson} quoted in
(\ref{s319},\ref{l:s324}), but in
flat disagreement with some of the more recent data listed in
(\ref{l:s324}). The double logarithms that occur at order $p^6$ in
the $K_{l3}$ form factors
have been determined recently \cite{kl3logs},  the full two-loop
calculation is
under way \cite{kl32loop}, and  the electromagnetic corrections
are  under investigation \cite{kl3electro}. A particular
combination of form factors of the vector currents has been studied at
two-loop order in \cite{post}.

\subsection{Future experiments}
The semileptonic $K_{l3}$ decays will be measured in the near future
at DAFNE \cite{kl3exp}. Of course, it will be very interesting to
compare the data with the prediction (\ref{s333}).

\section {Elastic $\pi\pi$-scattering}

The interplay between theoretical  and experimental
aspects of elastic $\pi\pi$ scattering
is illustrated in figure 1.
 On the theoretical side, Weinberg's calculation \cite{weinberg66}
of the scattering amplitude  at
leading order in the
low-energy expansion gives for the isospin zero S-wave scattering
 length the value
$a_{l=0}^{I=0}=0.16$ in units of the charged pion
mass. This  differs from the experimentally determined
value \cite{rosselet}
$a_0^0=0.26 \pm 0.05$ by two standard deviations. The one-loop
calculation \cite{glplb} enhances the leading order term to  $a_0^0=0.20\pm
0.01$ - the correction goes in the
right direction, but the result is still on the low side as far as the
 present experimental value is concerned.
 To decide about agreement/disagreement between theory and experiment,
 one should i)  evaluate
the scattering lengths in the theoretical framework at order $p^6$, and ii)
determine them more precisely experimentally.
Let me first comment on the theoretical work.
\begin{figure}[h]
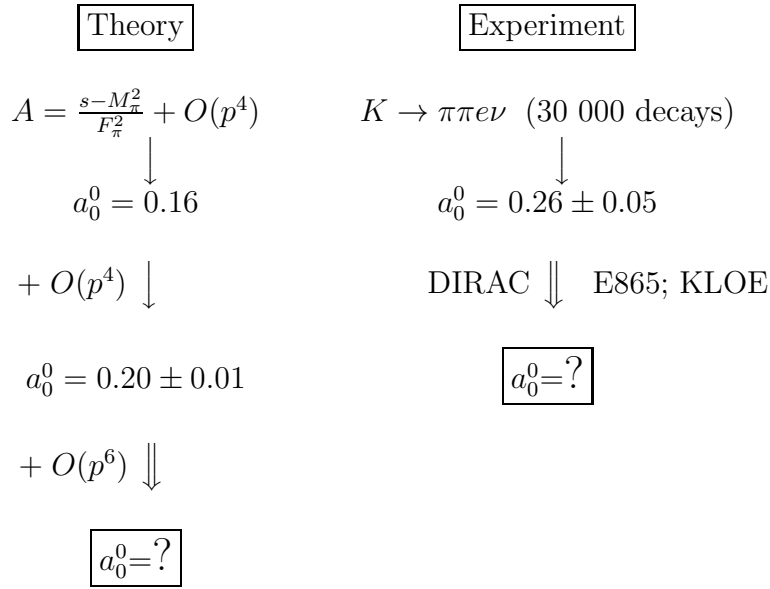

 \begin{center}
$
\hspace{1cm}\begin{array}{ccc}
\framebox{Theory}&\hspace{0cm}&\framebox{Experiment}
\\&&\\
A=\frac{s-M_\pi^2}{F_\pi^2}+O(p^4)&& K\rightarrow \pi\pi e \nu
\;\; (\mbox{30 000 decays})\\
\left.
\begin{array}{c} \\ \end{array}
\right\downarrow &&
\left.
\begin{array}{c} \\ \end{array}
\right\downarrow
\\
a_0^0=0.16 && a_0^0=0.26\pm 0.05
\\&&\\
\hspace{-1.5cm}\left.
\begin{array}{c}+\;O(p^4)  \\ \end{array}\right\downarrow&&
\hspace{1.3cm}\left.\begin{array}{c} {\mbox{DIRAC}} \\
\end{array}\right\Downarrow\begin{array}{c}
  {\mbox{ E865; KLOE}}\\ \end{array}
\\&&\\
a_0^0=0.20\pm 0.01&&\framebox{$a_0^0$=\mbox{\Large{?}}}
\\&&\\
\hspace{-1cm}\left.
\begin{array}{c}+\;O(p^6) \\ \end{array}\right\Downarrow
\begin{array}{c}  \\\end{array}&&
\\
\\
\framebox{$a_0^0$=\mbox{\Large{?}}}&&
\end{array}
$
\end{center}
 \caption{
Progress in the determination of the elastic $\pi\pi$ scattering
amplitude. References are provided in the text.}
\label{fig1}
\end{figure}

\subsection{ Theoretical aspects}

I consider QCD in the isospin symmetry limit
$m_u=m_d\neq 0$.
 Elastic $\pi\pi$ scattering
is then described by a single Lorentz invariant amplitude $A(s,t,u)$, that
depends on the standard
Mandelstam variables $s,t,u$.
 The effective lagrangian that describes
this process is given by a string of terms,
$
\calleff={\cal{L}}_2 +\hbar {\cal{L}}_4 +\hbar^2{\cal{L}}_6+\cdots\scs
$
where ${\cal{L}}_{n}$ contains $m_1$ derivatives of the pion fields and $m_2$
quark mass matrices, with $m_1+2m_2=n$  (here, I consider the standard
counting rules \cite{weinberg79,glannp}).
 The low-energy
expansion corresponds to an expansion of the scattering amplitude in
powers of $\hbar$,
\bea
A(s,t,u)=\left\{\begin{array}{ccccccc}\\
A_2&+&A_4&+&A_6& +&O(p^8)\co \\
\uparrow&&\uparrow&&\uparrow&&\\
{\mbox{tree}}&&{\mbox{1 loop}}&&{\mbox{2 loops}}&&\\
\end{array}\right.\label{eq3.1}
\eea
where  $A_n$ is of order $p^n$.
The tree-level result \cite{weinberg66} reads
\be
A_2=\frac{s-M_\pi^2}{F_\pi^2}\, ,
\ee
 and
the one-loop expression  $A_4$ may be found in \cite{glplb}. The
two-loop contribution $A_6$ was worked out in \cite{pipi6}.
(A dispersive evaluation of $A_6$ has been performed in
Ref.~\cite{knecht}
in the framework of generalized chiral perturbation theory, see below.
That  calculation is not sufficient for
the present purpose -
 what is needed for  the analysis outlined below is the
  complete two-loop expression of $A_6$ as presented in \cite{pipi6}.)

The amplitude $A_2+A_4+A_6$ contains several of the low-energy
constants
that occur in
${\calleff}$.
 In ${\cal{L}}_2$, there are two
 of them, the pion decay constant $F$ in the chiral limit, and the
parameter $B$, which are related to the condensate by
$
F^2B=-\langle 0|\bar{u}u|0\rangle$. In the loop expansion, these two
parameters can be expressed in terms of the
physical pion decay constant $F_\pi \simeq 92.4$ MeV and of the pion
mass, $M_\pi = 139.57$ MeV.
The $\pi\pi$ scattering amplitude contains, in the two-loop approximation,
 in addition several LEC's
occurring in ${\cal{L}}_{4}$ and in ${\cal{L}}_6$,
\bea
\begin{array}{ll}
\left.\begin{array}{l}
 {\cal{L}}_2:F_\pi, M_\pi\\
 {\cal{L}}_4:\bar{l}_1,\bar{l}_2,\bar{l}_3,\bar{l}_4\\
 {\cal{L}}_6: \bar{r}_1,\ldots, \bar{r}_6\end{array}\right\}&{\mbox{occur in}}
\;\pi\pi\rightarrow\pi\pi\; {\mbox{ (two-loop approximation)}}.\end{array}
\label{eq2.1}
 \eea
These LEC's are not determined by chiral symmetry - they are,
 however, in
principle calculable in QCD \cite{rebbi}.

Once the amplitude is available in algebraic form, it is a trivial
matter to  evaluate  the threshold
parameters. To quote an example, the isospin zero
S-wave scattering length is of the form
\bea
a_0^0&=&\frac{7M_\pi^2}{32\pi F_\pi^2}\left\{1+c_4 x +c_6 x^2
+O(p^8)\right\}\, ; \,
x=\frac{M_\pi^2}{16\pi^2F_\pi^2}\, .
\eea
The coefficients  $c_4,c_6$ contain
the low-energy constants listed in (\ref{eq2.1}).
Similar formulae hold for all other threshold parameters - the
explicit expressions  for the scattering lengths and effective ranges of the
S-and P-waves as well as for the D-wave scattering lengths at order
$p^6$ may be found
in \cite{pipi6}. It is clear that, before a numerical value for these
parameters can be given, one needs an estimate of the low-energy
constants. The calculation is under way - it is, however, quite
involved: One has to solve numerically the Roy-equations \cite{roy} with input
from the high-energy absorptive part. Second, one assumes that the
couplings that describe the mass dependence of the amplitude may be
estimated from resonance exchange. Requiring that the experimental
amplitude agrees near threshold with the chiral representation allows
one finally to pin down the remaining couplings, as well as the
scattering lengths $a_0^0$ and $a_0^2$. The remaining threshold
parameters may then be obtained from the Wanders sum rules \cite{wanders}.
The first part of the program is completed, and the report will
appear soon \cite{acgl}. The second
part, that will allow us to predict the values of all threshold
parameters, is under investigation \cite{pipi6a}.

\subsection{Threshold parameters from experimental data}

On the {\it experimental} side, several  attempts are under way to
 improve our knowledge of the threshold parameters.
 The most
promising ones among them are i) semileptonic $K_{l4}$ decays with improved
 statistics, E865 \cite{e865} and KLOE \cite{kloe},
 and ii) the   measurement of the pionium lifetime -
 DIRAC \cite{dirac} - that will allow one to directly determine
 the combination $|a_0^0-a_0^2|$ of
 S-wave scattering lengths.
It was one of the aims of last years workshop in Dubna \cite{dubna}
 to discuss  the precise relation between
the lifetime of the pionium atom and the  $\pi\pi$ scattering
lengths -
  I refer the
interested reader to the numerous  contributions to that workshop
for details.
 Let me note that recently,
 using the effective lagrangian framework proposed by Caswell and Lepage some
 time ago \cite{caswell}, the width of pionium in its ground state has
 been determined \cite{gglr} at leading and next-to-leading order in isospin
 breaking and to all orders in the chiral expansion.
  This result will
 allow one to evaluate the combination $|a_0^0-a_0^2|$ with high
 precision, provided that DIRAC determines the lifetime at the 10\%
 level, as is foreseen \cite{dirac}.
\subsection{Why do we wish to know  the scattering lengths?}
Why are we  interested in a precise value of the scattering length
$a_0^0$? First, it is one of the few occasions that
 a quantity in QCD
 can be predicted rather precisely - which is, of course, by itself worth
checking. Second, as has been pointed out in
 \cite{gchpt}, this prediction assumes that the condensate has the
standard size in the chiral limit - in particular, it is assumed to
be non vanishing. For this reason, the authors of Ref.~\cite{gchpt} have
reversed the argument and have set up a framework where the condensate
is allowed to be small or even vanishing in the chiral limit - the
so called generalized chiral perturbation theory
\footnote{Let me note that there is
no sign for a small
condensate in present lattice calculations \cite{lattice}.
Further interesting investigations of the small condensate  scenario
 have  been
performed in Refs. \cite{derafael,kogan}.}.
 Whereas the S-wave
scattering lengths cannot be predicted in that framework, one may
relate their size to the value of  the condensate. Hence, measuring
 $a_0^0$,  $a_0^2$ or a combination thereof \cite{dirac}
 may allow one to determine the
nature of chiral
symmetry breaking by experiment \cite{gchpt,pipigchpt}.

\section{Conclusion}
Chiral perturbation theory has a wide field of
applications. Many of its predictions  have already been tested
 \cite{handbook,mainz,eta}, and many more will be investigated in the
near future,
 e.g.   by E865 \cite{e865} in Brookhaven,
 by DIRAC \cite{dirac} at CERN, and by DAFNE in Frascati \cite{handbook}.

\section*{Acknowledgements}
It is  a great pleasure to thank the organizers
for the most interesting and enjoyable stay at the Budker Institute
 of Nuclear Physics in Novosibirsk.

\end{document}